\documentclass[12pt]{article}
\usepackage{epsfig}

\usepackage{latexsym, graphicx}
\newcommand{\be}{\begin{equation}}
\newcommand{\ee}{\end{equation}}
\newcommand{\bea}{\begin{eqnarray}}
\newcommand{\eea}{\end{eqnarray}}

\begin{document}

\title{\vbox{\baselineskip15pt
\hfill \hbox{\normalsize
{hep-ph/0510081}} \\
\hfill \hbox{\normalsize {EFI-2005-17}} } Models of Baryogenesis
via
\\ Spontaneous Lorentz Violation}

\author{{Sean M.
Carroll\thanks{carroll@theory.uchicago.edu} \ and Jing Shu\thanks{jshu@theory.uchicago.edu}}\\
\normalsize{Enrico Fermi Institute, Deptartment of Physics,}\\
\normalsize{and Kavli Institute for Cosmological Physics} \\
\normalsize{University of Chicago, 5640 S. Ellis Avenue}\\
\normalsize{Chicago, IL 60637, USA} }

\maketitle

\begin{abstract}

In the presence of background fields that spontaneously violate
Lorentz invariance, a matter-antimatter asymmetry can be generated
even in thermal equilibrium.  In this paper we systematically
investigate models of this type, showing that either high-energy
or electroweak versions of baryogenesis are possible, depending on
the dynamics of the Lorentz-violating fields.  In addition to the
previously-studied models of spontaneous baryogenesis and
quintessential baryogenesis, we identify two scenarios of
interest: baryogenesis from a weak-scale pseudo-Nambu-Goldstone
boson with intermediate-scale baryon-number violation, and
sphaleron-induced baryogenesis driven by a constant-magnitude
vector with a late-time phase transition.
\end{abstract}

\vfill\eject

\parindent=20pt
\baselineskip=14pt

\section{Introduction}

The observed universe manifests a pronounced asymmetry between
the number density of baryons $n_b$ and antibaryons $n_{\bar b}$
(see {\it e.g.} \cite{rad-ann}).   Numerically, the
baryon-to-entropy ratio is
\begin{equation}\label{eta_B} \eta_{B}
\equiv \frac{n_{B}}{n_{\gamma}} = 9.2^{+0.6}_{-0.4} \times
10^{-11} \, ,
\end{equation}
where $n_{B} = n_{b} - n_{\bar{b}}$ and $n_{\gamma}$ are the
baryon and photon number density, respectively. However, the origin
of the baryon number asymmetry remains a major puzzle for cosmology
and particle physics.

In a classic work, Sakharov argued that three conditions are
necessary to dynamically generate a baryon asymmetry in an
initially baryon-symmetric universe: (1) baryon number
non-conserving interactions; (2) C and CP violation; (3) departure
from thermal equilibrium \cite{SP67,BG}.  In deriving these
conditions, however, the assumption is made that CPT is conserved,
as it will always be in a Lorentz-invariant local quantum field
theory.  If Lorentz invariance is violated, CPT may also be
violated, and the assumption of departure from thermal equilibrium
is no longer necessary
\cite{Dolgov:1981hv,BCKP96,Carmona:2004xc,DiGrezia:2005yx}. A
concrete implementation of this idea is given by the ``spontaneous
baryogenesis'' \cite{CK88} scenario, which has subsequently been
elaborated upon in various ways \cite{BCKP96, Cohen:1991iu,
Trodden, XMZ, Yamag, Chiba:2003vp, Alberghi:2003ws, DKKMS04}.

Lorentz invariance is spontaneously violated whenever a tensor
field has a nonzero expectation value. In recent years there has
been considerable interest in this possibility in the context of
various quantum field theories, extra dimensions and brane-world
scenarios as well as modified gravity and string theories
\cite{KS89, Mad99, Carroll:2001ws, OV03, Bertolami:2003nm, CFN01,
Bjo01, LFL03, Jen04, CFMN04, ACLM03, Gri04, JLM04, MS01, AC02}.
The existence of Lorentz violation leads to interesting
implications for neutrino experiments \cite{Kostelecky:2003cr},
high energy cosmic ray phenomena \cite{KS89, CG97,
Bertolami:2000qa}, evolution of the fine-structure constant
\cite{KLP03,Bertolami:2003qs} and Newton's constant \cite{CL04},
the cosmological constant problem \cite{KT02, Bertolami:1997iy},
dark energy \cite{ACLM03, Bertolami:2003qs, DeDeo04}, inflation
\cite{ACMZ03}, and the cosmic microwave background (CMB)
\cite{Lim:2004js}.

In this paper we consider a vector field $A_\mu$ with a timelike
expectation value in a cosmological background.  The norm of the
vector need not be constant, nor need the vector field be fundamental;
we consider different possible dynamical origins for the field,
including the possibility that it is the gradient of a scalar.
A coupling to the baryon-number current $J^\mu_B$ of the form
\be
  {\cal L} = gA_\mu J^\mu_B
\ee
leads to an effective chemical potential for baryons of the form
$\mu = -gA_0$.  In such a background, baryons and antibaryons have
different masses, and
a nonzero value of $n_B$ will arise
even in thermal equilibrium.  Depending on the behavior of
$\mu/T$, where $T$ is the temperature of the cosmological plasma,
a relic baryon asymmetry can be generated by interactions that
violate either $B+L$ or $B-L$, as we explore below.

We identify two scenarios of potential interest.  One is the case
of a simple constant-magnitude timelike vector field coupled to
$J^\mu_{B+L}$.  In this case we calculate that there can be an
appropriate baryon asymmetry generated by electroweak sphalerons
\textsl{alone}.  However, the required magnitude for the vector is
in conflict with bounds from present-day experiments, so it is
necessary to invoke a late-time phase transition to eliminate the
vector today. The other possibility is that of a
derivatively-coupled pseudo-Nambu-Goldstone boson.  We find that
the PNGB mass parameter required to generate the correct asymmetry
is naturally at the weak scale.  This scenario requires
($B-L$)-violating interactions that freeze out at an intermediate
scale of order $10^{10}$~GeV, which is perfectly reasonable in
models of Majorana neutrino masses.  We believe that this model is
worthy of further study.

The outline of this paper is as follows.  The next section
describes how a baryon asymmetry will arise in the presence of
spontaneous Lorentz violation.  We pay particular attention to the
calculation of the relevant freeze-out temperature, which depends
on the sources of baryon-number violation as well as the dynamics
of the background vector field, and we discuss some specific
sources of baryon-number violation.  A case of special interest
is the effect of sphalerons, which often tend to dilute any existing
baryon asymmetry; in Section~\ref{sphalerons} we show that sphalerons
can actually be responsible for the observed baryon asymmetry in
the presence of an appropriate Lorentz-violating background.
In Section \ref{constraints} we
consider some indirect and direct experimental constraints on the
vector field $A_\mu$ at late
times.  These bounds are much lower than the required value
to generate the right net baryon number density, implying that
the vector must decay appreciably between the early universe and
today.  We then consider models for the Lorentz-violating fields
themselves, including both models with a constant vector
field (either fundamental or a ghost condensate), and
models featuring scalar fields rolling down a potential.

\section{Baryogenesis in the presence of Lorentz violation}
\label{setup}

\subsection{Basic mechanism}
\label{basic}

We consider the theory of a vector field $A_\mu$ with a nonzero
vacuum expectation value (vev), coupled to a current $J^\mu$.  The
action is given by
\begin{equation}\label{Eq2-1}
\mathcal{S} = \int d^{4}x  \sqrt{-g}( \mathcal{L}_{A} + \mathcal{L}_{int} +
\mathcal{L}_{m})
 \, ,
\end{equation}
where $\mathcal{L}_{A}$ is the Lagrange density of the vector
field, $\mathcal{L}_{m}$ denotes the Lagrange density for the
other matter fields and $\mathcal{L}_{int}$ is the Lagrange
density for the interaction term. The vector $A_{\mu}$ is not
necessarily the field that is varied to get the equations of
motion from this action; it could be the derivative of a scalar
field $\partial_{\mu} \phi$ \cite{CK88, Cohen:1991iu, Trodden,
XMZ, Yamag, Chiba:2003vp, ACLM03}, the vector current of some
hidden fermions $\bar{\psi} \gamma^{\mu} \psi$ \cite{Bjo63, CFN01,
Bjo01, LFL03, Jen04}, the four-divergence of some higher-rank
tensor background field $\nabla_{\mu} \nabla_{\nu} \ldots
T^{\mu\nu \cdots}$ \cite{BCKP96, KS89}, or even the derivative of
the scalar curvature $\partial_{\mu} \mathcal{R}$ \cite{DKKMS04}.

Our concern will be with the effect of the vev for $A_\mu$, so we
will not investigate possible forms of the vector kinetic term until
Section~\ref{lorentz}. The crucial feature will be the existence
of a timelike expectation value, \be
  \langle A_\mu A^\mu\rangle < 0\,.
\ee
In a flat Robertson-Walker universe with metric
\be
  ds^2 = -dt^2 + a^2(t)[dx^2 + dy^2 + dz^2]\ ,
\ee
we will take the vector to point purely in the timelike
direction, with components
\be
  A_\mu = (a_0, 0,0,0)\ .
  \label{config}
\ee Note that $a_0$ is {\em not} assumed to be constant.
Condensation of the vector field clearly leads to spontaneous
breaking of Lorentz invariance; we are assuming that the rest
frame of the vector field is that of the cosmological fluid. Such
a vector field will, of course, have an energy-momentum tensor
with a corresponding effect on the expansion rate; for the
purposes of this paper, however, we will assume that this energy
remains negligible compared to the total energy in the
radiation-dominated era.

We take the
interaction Lagrangian density to be the natural one\footnote{The
interaction term here can not be eliminated by a field
redefinition because the conservation of the current $J_{\mu}$ is
violated by the matter fields, such violation (for example, baryon
current violation) is essential to generate the matter-antimatter
asymmetry in the universe.}
\begin{equation}\label{Eq-2}
\mathcal{L}_{int} = g A_{\mu} J^{\mu}
 \, ,
\end{equation}
where $g$ is a coupling constant and $J^{\mu}$ is the current
corresponding to some continuous global symmetry of the matter
fields such as baryon number.  This current takes the form
\be
  J^\mu = \sum_i \beta_i \bar{\psi_i}\gamma^\mu\psi_i
\ee
for some set of dimensionless parameters $\beta_i$.
For the configuration (\ref{config}), the interaction becomes
\begin{equation}\label{Eq-1}
\mathcal{L}_{int} = - g a_{0} Q
 \, ,
\end{equation}
where $Q$ is the conserved charge density.

For purposes of illustration, consider the case when
$J^{\mu}$ is precisely the baryon number current.  Eq. (\ref{Eq-1})
then becomes
\be
  \mathcal{L}_{int} =- g a_{0} (n_{b} - n_{\bar{b}})\,,
\ee where $n_b$ is the number density of baryons and $n_{\bar b}$
that of antibaryons.  The vector-field background spontaneously
violates CPT invariance, splitting the masses of the baryons and
antibaryons. In the Appendix we verify that this has the effect of
giving baryons a chemical potential \be \mu^0_{b} = g a_{0}\,, \ee
with antibaryons getting a corresponding effective chemical
potential $\mu^0_{\bar{b}} = - g a_{0}$. Because of this effect,
in thermal equilibrium there will be a non-zero baryon number
density generated by baryon-number violating interactions:
\begin{equation}\label{Eq-3}
n_{B} = n_{b} - n_{\bar{b}} = \frac{g_{b}T^{3}}{6 \pi^{2}} \Big[
\pi^{2} \frac{\mu_{b}^0}{T} + \Big(\frac{\mu_{b}^0}{T} \Big) ^{3}
\Big] \simeq \frac{g_{b}\mu_{b}^0 T^{2}}{6} \sim \mu_{b}^{0} T^{2}
 \, ,
\end{equation}
where $g_{b}$ counts the internal degrees of freedom of the baryons.
In thermal equilibrium, the entropy density is
$s = (2\pi^{2}/45) g_{\ast s}
T^{3}$. When the $B$-violating interactions mentioned above become
ineffective ($\Gamma \leq H$), we get the final baryon asymmetry
\begin{equation}\label{Eq-4}
{n_{B} \over s} \sim \frac{\mu_b^0}{g_{\ast s}T_{F}} = \frac{g
a_0}{g_{\ast s}T_{F}}
 \, ,
\end{equation}
where $T_{F}$ is the temperature at which the baryon number
production is frozen out.

This is the effect that we will explore in this paper:  the
generation of an {\sl equilibrium} baryon asymmetry due to
spontaneous Lorentz violation.  This phenomenon has been
considered previously in various specific contexts \cite{CK88,
BCKP96, Trodden, XMZ, Yamag,DKKMS04}; here we consider the
scenario in some generality.  Our concern is therefore to
determine what happens for different sources of baryon-number
violation (and hence freeze-out temperatures) and Lorentz
violation (and hence chemical potentials).

\subsection{Freeze-out temperature}
\label{freeze-out}

Our first step is to consider possible values of the freeze-out
temperature $T_F$ for different sources of baryon-number
violation. We know that sphaleron transitions \cite{Spha} connect
baryon number and lepton number, so we need to consider both the
baryon number current and lepton number current that couple to the
background fields\footnote{More generically, not only the
lepton current, but any currents derivatively coupled to the scalar
field that are non-orthogonal to the baryon number current should
be considered, as their relaxation energetically favors a non-zero
baryon charge. This is similar to the case of spontaneous
electroweak baryogenesis \cite{Cohen:1991iu}.}. The interaction
Lagrangian will actually be a sum of couplings to the baryon and
lepton currents,
\begin{equation}\label{E15a}
  \mathcal{L}_{int} = g_{B} A_{\mu} J_{B}^{\mu} + g_{L} A_{\mu}
  J_{L}^{\mu} \ .
\end{equation}
It is convenient to rewrite this interaction in terms of the
$B+L$ and $B-L$ currents,
\begin{equation}\label{E15}
  \mathcal{L}_{int} =  g_{-} A_{\mu} J_{B-L}^{\mu} + g_{+}
  A_{\mu} J_{B+L}^{\mu}
 \, ,
\end{equation}
where
\be
  J_{B-L} = J_{B}- J_{L} , \quad J_{B+L} = J_{B}+ J_{L} ,
\ee
and
\be
  g_{-} = {1\over 2}(g_{B} - g_{L}) , \quad g_{+} = {1\over 2}(g_{B} + g_{L}) .
\ee

Sphaleron transitions will induce a nonzero baryon number
density if the Lorentz-violating background is coupled to the
$J_{B+L}^\mu$ current; that is, if $g_+$ is nonvanishing.
{}From Eq. (\ref{Eq-4}), we know that
\begin{equation}\label{E16}
\frac{n_{B-L}}{s} = \frac{\mu_{-}^0(T_{-})}{g_{* s} T_{-}} =
\frac{g_{-} a_0(T_{-})}{g_{* s} T_{-}}\ ,
\qquad \frac{n_{B+L}}{s} = \frac{\mu_{+}^0(T_{+})}{g_{* s}
T_{+}}= \frac{g_{+} a_0(T_{+})}{g_{* s} T_{+}}
 \, ,
\end{equation}
where $T_{-}$ and $T_{+}$ are the lowest freeze-out temperature
for any interactions that could violate $B-L$ and $B+L$,
respectively. The ratio ${n_{B+L}}/{s}$ arises from
sphaleron transitions, so we know that $T_{+}$ must be
150~GeV, which is the critical temperature of electroweak phase
transition when sphalerons freeze out. If there is no
additional symmetry to set $g_{-} \gg g_{+}$ or $g_{-} \gg g_{+}$,
we will assume that $g_{-}$ and $g_{+}$ are of the same order. We
also presume that $T_{+}= 150~\textrm{GeV} \ll T_{-}$, which is at
least in the order of TeV, so whether $n_{B+L} \ll n_{B-L}$ or $n_{B+L}
\gg n_{B-L}$ will only depend on whether $\mu^0(T)/T$ is an
increasing or decreasing function with respect to $1/T$.  (We think
of $\mu^0(T)/T$ as a function of $1/T$, rather than $T$, since the
former is evolves monotonically with the scale factor $a$.)
\begin{figure}
\centerline{
\psfig{figure=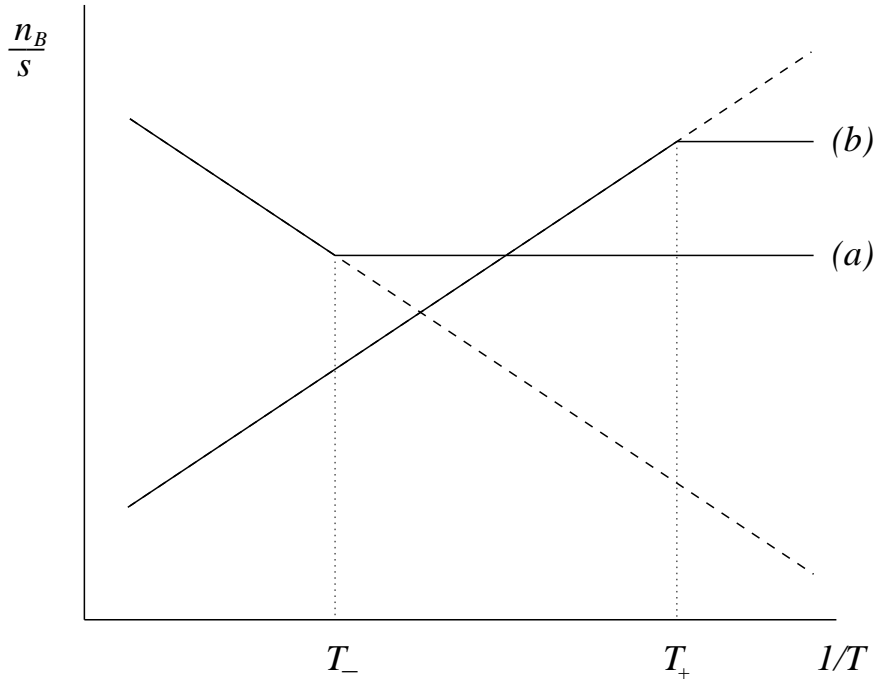,height=9cm}}
  \caption{Two cases for the evolution of the baryon-to-entropy ratio
  in the presence of a Lorentz-violating effective chemical potential
  $\mu^0$.  In case $(a)$, $\mu^0/T$ increases or remains constant
  as a function of $1/T$ (that is, as
  the universe expands), and the final baryon asymmetry is determined
  by the temperature $T_+$ at which $(B+L)$-violating interactions
  freeze out.  Since electroweak sphalerons violate $(B+L)$, this
  will be at $T_+ \approx 150$~GeV.
  In case $(b)$, $\mu^0/T$ decreases as a function of
  $1/T$, and the final baryon asymmetry is determined
  by the temperature $T_-$ at which $(B-L)$-violating interactions
  freeze out, presumably at some higher temperature.}
  \label{nbosfig}
\end{figure}
We therefore have
\begin{eqnarray}\label{E16a}
\begin{array}{l}
n_{B+L} \gg n_{B-L} \, ~~~ \displaystyle \textrm{if~}
\frac{|\mu^0(T)|}{T} ~~
\textrm{increases as a function of $1/T$,} \\
n_{B+L} \ll n_{B-L} \, ~~~ \displaystyle \textrm{if~}
 \frac{|\mu^0(T)|}{T} ~~
\textrm{decreases as a function of $1/T$.}
\end{array}
\end{eqnarray}
The net baryon number $n_{B} = (n_{B+L} + n_{B-L})/2$, so we know
that $n_{B}$ is of the same order as max$\{n_{B+L}$, $n_{B-L}\}$.
{}From Eq.~(\ref{E16a}), we get
\begin{eqnarray}\label{E16b}
  \frac{n_B}{s} &=& \left\{
  \begin{array}{l}
  \displaystyle \frac{n_{B+L}}{2s} \sim \frac{g_{+}
  a_0(T_{+})}{g_{* s} T_{+}}
  \, ~~ \textrm{if~}
  \frac{|\mu^0(T)|}{T} ~~
  \textrm{increases as a function of $1/T$,} \\
  \displaystyle \frac{n_{B-L}}{2s} \sim \frac{g_{-}
  a_0(T_{-})}{g_{* s} T_{-}}  \, ~~  \textrm{if~}
  \frac{|\mu^0(T)|}{T} ~~
  \textrm{decreases as a function of $1/T$.}
  \end{array}\right.
\end{eqnarray}
Thus, the behavior of the effective chemical potential $\mu^0$
as a function of temperature controls the final baryon asymmetry
by determining whether the freeze-out temperature $T_{F}$ is
$T_{+}$ or $T_{-}$.  Figure~\ref{nbosfig} illustrates the two
basic possibilities.

The interesting thing in our paper is the former case.
That is, if $a_0(T)$ is a constant or even increases
with time (or decreases more slowly than $T$),
we will get $n_{B+L}/ s \gg n_{B-L}/ s$, so that $n_{B}$
will only depend on the sphaleron freeze-out temperature
$T_{+}=150 $~GeV.  Unfortunately, the corresponding value for
$\mu^0$ is naively in conflict with present-day experimental limits,
as we discuss in Section~\ref{constraints}.  In Section~\ref{lorentz},
we consider various explicit scenarios for the origin of the
Lorentz-violating field, and mechanisms by which it may evade
experimental constraints.

The latter case, where $a_0(T)$ decreases faster than $T$,
will apply if the interaction arises from
higher-power derivatives in the current or a slowly-rolling
field.  The net baryon number generated
from sphaleron transitions may then be neglected if there
is any higher-temperature $B-L$ violation, since
$n_{B+L}/ s \ll n_{B-L}/ s$ and $n_{B}$ will only depend on
$T_{-}$.

In a nonsupersymmetric model, the most natural way to violate
$B-L$ is to introduce a Majorana mass term which violates $L$ by
two units. Following the current mass boundary of light left-handed
neutrinos from the atmospheric and solar neutrino oscillation
experiments \cite{neutrino, Moh04} and the analysis of WMAP
\cite{WMAP} and SDSS \cite{SDSS}, the freeze-out temperature
$T_{-}$ is $10^{13}$~GeV for light neutrinos with hierarchical
masses and $10^{11}$~GeV for light neutrinos with degenerate masses.
These limits arise from
considering scattering process mediated by heavy right-handed
neutrinos \cite{XMZ}. In a supersymmetric model, the $L$-violating
interactions (and therefore the $B-L$ violation interactions) are not
only from the Majorana mass term, but also from the mass of its
scalar partners. We consider the case in which the right-handed
neutrinos and their SUSY partner sneutrinos have the mass of order
TeV \cite{AHMSW01, BN01}. Then we obtain $T_{-} \approx 1 $~TeV
since the decay processes will freeze out at a temperature near
their mass scale.  Note that low-energy baryogenesis offers a way
out of the gravitino problem of supersymmetric inflation, since
the reheat temperature can be low enough not to overproduce the
gravitinos.

\section{Baryogenesis via sphalerons}
\label{sphalerons}

Baryon-number violation in the Standard Model arises from
sphaleron transitions. In this section we consider carefully
the baryon asymmetry generated by sphalerons in the presence of
the effective chemical potential $\mu$ induced by Lorentz violation.
Our analysis closely follows that of \cite{BCKP96}, with one
important exception:  we keep the effective chemical potential
induced by (\ref{Eq-2}) in the expression for the free
energy, as derived in the Appendix.  This opens the possibility
of generating the baryon asymmetry from sphalerons without any
additional high-energy baryon violation.

Consider $N$ generations of quarks with mass $m_{q_{i}}$ and leptons
with mass $m_{l_{i}}$, for $i = 1, \ldots , N$. The free energy in a
unit volume for the system in equilibrium at temperature $T$ is
given by
\begin{equation}\label{Eq3-2}
\mathcal{F} = 6 \sum_{i=1}^{2N} F(m_{q_{i}}, \mu) + \sum_{i=1}^{N}
[2F(m_{l_{i}}, \mu_{i}) + F(0, \mu_{i})] \, ,
\end{equation}
where the parameters $\mu$ and $\mu_{i}$ are the chemical
potentials of the quarks and the $i$th lepton, respectively. We do
{\em not} include the Lorentz-violating contribution $\mu^0$ in
the chemical potential, but instead include it explicitly in the
expression below for the particle energy.  The
chemical potentials for the leptons in the same doublet will
be equal, since SU(2) interactions are in equilibrium. We neglect
the neutrino masses as they are too small to have an important
effect. We also do not take the
right-handed neutrinos into account in the free energy density; either
the right-handed neutrinos have decayed out of equilibrium by the time
sphalerons enter into thermal equilibrium, or the right-handed neutrino
is simply ``sterile" so that it is decoupled from the electroweak
anomaly.

In the expression (\ref{Eq3-2}), the free energy density
for a fermion of mass $m$ and chemical potential $\mu$ is given
by
\begin{equation}\label{Eq3-3}
F(m, \mu) = - T \int \frac{d^{3}K}{(2 \pi)^3} [\ln(1+
e^{-(E+\mu)/T}) + \ln(1+ e^{-(\bar{E} - \mu)/T})] \, ,
\end{equation}
where $K_i$ is the momentum of the fermion, $E= \sqrt{K_i^2 + m^2}
+ \mu^{0}$ is the energy of the fermion and $\bar{E}= \sqrt{K_i^2 +
m^2} - \mu^{0}$ is the energy of antifermion.\footnote{For the
general case, if we need to consider the isocurvature effect of
the vector background (see Ref. \cite{TCK90} for such effect in
spontaneous baryogenesis), we cannot find a preferred
rest frame so that the vector field is purely timelike
everywhere. In the expression of the local free energy
$F(\vec{x}, m, \mu)$, the energy of the fermion is
$E= \sqrt{(\vec{k}-g \tilde{Q}_{f} \vec{A})^2 + m^2} + \mu^{0}$
according to Eq. (\ref{Eq2-6}). The reason we use the expression $E=
\sqrt{(\vec{k}-g \tilde{Q}_{f} \vec{A})^2 + m^2} + \mu^{0}$
instead of $E= \sqrt{{{k}}^2 + m^2} + \mu^{0}$ is that we
integrate over the canonical conjugate momentum $\vec{P}$ in the
phase space of the microcanonical ensemble.}
Here we choose to
treat the Lorentz-violating interaction $\mu^0$ as a contribution
to the energy of each particle (opposite for baryons and
antibaryons), rather than as a contribution to the chemical
potential; the two choices are completely equivalent.
In Ref. \cite{BCKP96}, the the additional contribution $\mu^{0}$
in the fermion energy was neglected; in that case, sphalerons
can dilute the baryon density but not produce it.

For quarks and each species of leptons, the $\mu^{0}$ term is
different due to the different coupling strength of the
interaction between vector background and the corresponding fermi
current. The existence of such a constant energy difference simply
shifts the energy of the fermion and antifermion. We use $\mu^{0}$
and $\mu^{0}_{i}$ to denote such energy shift for quarks and the $i$th
species of leptons respectively. The expression for leptonic and
baryonic number densities are
\begin{equation}\label{Eq3-4}
l_{i} = \frac{d}{d \mu_{i}} [2F(m_{i}, \mu_{i}) + F(0, \mu_{i})]
\,
\end{equation}
and
\begin{equation}\label{Eq3-5}
B = 2 \frac{d}{d \mu} \sum_{i=1}^{2N} [F(m_{q_{i}}, \mu)] \, .
\end{equation}
(Note that $B$ and $n_b$ are the same quantity.)
In the high-temperature approximation $m^{2}/T^{2} \ll 1$, we get
\begin{equation}\label{Eq3-6}
F(m, \mu) \approx F(m, 0) - \frac{1}{12} (\mu + \mu^{0})^{2} T^{2}
(1 - \frac{3}{2 \pi^2} \frac{m^2}{T^2}) \, .
\end{equation}
Sphaleron transitions violate $l_i$ and $B$, but preserve the
combinations
\be
  L_i = l_i - N^{-1}B\, .
\ee
The conserved number densities $L_{i}$ are then given by
\begin{equation}\label{Eq3-7}
L_{i} \equiv l_{i} - N^{-1} B \approx \frac{(\mu + \mu^0) T^2}{3N}
\alpha - \frac{(\mu_{i} +\mu_{i}^{0}) T^2 }{2} \beta_{i}  \, ,
\end{equation}
where
\begin{equation}\label{Eq3-8}
\alpha \equiv 2N - \frac{3}{2 \pi^2} \sum_{i=1}^{2N}
\frac{m_{q_{i}}^2}{T^2}, ~~~~~ \beta_{i} \equiv 1 -
\frac{1}{\pi^2} \frac{m_{l_{i}}^2}{T^2} \, .
\end{equation}

Effectively, the sphaleron transitions convert $3N$ quarks and 1
lepton into nothing. In thermal equilibrium, this
leads to the relation $\mu = - \sum_{i}^{N} \mu_{i} / 3N$. We
solve $\mu_{i}$ from Eq.(\ref{Eq3-7}) and summing over $i$ leads to
the expression
\begin{equation}\label{Eq3-9}
\mu = \Big( \frac{2}{T^2}\sum_{i=1}^{N} \frac{L_{i}}{\beta_{i}} +
\mu^{0}_{\Delta} \Big) \Big( \frac{2}{3N} \sum_{i=1}^{N}
\frac{\alpha}{\beta_{i}} + 3N \Big)^{-1}  \, ,
\end{equation}
where
\begin{equation}\label{Eq3-10}
\mu^{0}_{\Delta} = \sum_{i=1}^{N} \mu_{i}^{0} - \frac{2\alpha}{3N}
\mu^{0} \sum_{i=1}^{N} \frac{1}{\beta_{i}} \, .
\end{equation}
{}From the baryonic density Eq.~(\ref{Eq3-5}), we get
\begin{equation}\label{Eq3-11}
B = - \frac{1}{3} (\mu + \mu^{0}) T^2 \alpha\, .
\end{equation}
Plugging in $\mu$, the final baryon number density can be written
\be
  B = B^{(0)} + B^{(\mu)}\, ,
\ee
with
\begin{equation}\label{Eq3-12}
B^{(0)} = - 2\alpha \Big( \sum_{i=1}^{N} \frac{L_{i}}{\beta_{i}} \Big)
\Big( 9N + \frac{2\alpha}{N} \sum_{j=1}^{N}
\frac{1}{\beta_{i}} \Big)^{-1}
\ee
and
\be
B^{(\mu)} =  \frac{1}{3}\alpha T^2 \Big[ \mu^0 + 3\mu^{0}_{\Delta} \Big(
\frac{2 \alpha}{N} \sum_{j=1}^{N} \frac{1}{\beta_{j}} + 9N
\Big)^{-1} \Big]  \, .
\end{equation}
$B^{(0)}$ is the conventional baryon number in the presence of
sphalerons, which can be expressed as
\begin{equation}\label{Eq3-13}
B^{(0)} = \left\{
\begin{array}{l}
\displaystyle - \frac{4}{13 \pi^2}\sum_{i=1}^N L_i \frac{m_{l_i}^2}{T^2} \, ~~~ B - L =0 \\
\frac{4}{13} (B -L) \, ~~~ B - L \neq 0\, .
\end{array}\right.
\end{equation}
This is the result obtained in Ref.~\cite{KRS87}. If the
initial $B-L$ is nonzero (usually a nonzero $L$) when sphalerons
enter into thermal equilibrium, then sphaleron transitions will
not wash out the initial $B$ asymmetry and will convert an $L$
asymmetry into $B$ asymmetry in some cases. If the initial $B-L$
is zero, taking the leptoquark decay to be dominated by the
heaviest lepton ${\tau}$ and the free-out temperature at the
electroweak scale, we will get a dilution by a factor of about
$10^{-6}$ \cite{KRS87, KRS85}.

For the term generated by the Lorentz-violating interaction,
at leading order $\alpha \sim 2N$, and
$\beta_{i} \sim 1$, and we obtain
\begin{equation}\label{Eq3-14}
B^{(\mu)} = - \frac{2N}{13} T^2 (3 \mu^{0} + \frac{1}{N} \sum_{i=1}^{N}
\mu_{i}^{0}) \, .
\end{equation}
If there is leptonic flavor violation in thermal equilibrium, we
will consider the vector background coupled to the lepton current
directly, so the sum over all the species of the effective
chemical potential will become a single one $\mu_{L}^{0} \equiv
\displaystyle \frac{1}{N} \sum_{i=1}^{N} \mu_{i}^{0}$. Since each
quark carries baryon number 1/3, the effective chemical
potential for quarks is $\mu_{B}^{0} = 3\mu^{0}$. So
Eq.~(\ref{Eq3-13}) becomes $B \propto (\mu^{0}_{B} + \mu^{0}_{L}) =
2 \mu^{0}_{B+L}$.

We therefore see that the baryon asymmetry in the presence of
the interaction (\ref{Eq-2}) is proportional to $\mu^0T^2$, just
as in (\ref{Eq-3}).  Thus, a
nonzero net baryon number density can in principle be spontaneously
generated through sphaleron transitions in thermal equilibrium
in the presence of a
nonzero time-like vector background coupled to $J_{B+L}$
current.

\section{Present-day constraints on Lorentz violation}
\label{constraints}

We have calculated the baryon-to-entropy ratio $n_B/s$ as a function
of the coupling $g$, the vector field magnitude $a_0$, and the freeze-out
temperature $T_{F}$, and argued that low-temperature baryogenesis is
possible for sufficiently large $\mu^0=ga_0$.  However, we must take into
account the experimental constraints on this parameter in the
present-day universe.  Although the constraints are not airtight,
we find that they are in apparent conflict with
the values of $ga_0$ required for low-temperature baryogenesis in
the absence of fine-tuning.
It is therefore necessary to consider models in which $a_0$ decays
between early times and today.

\subsection{Direct Constraints from mesons}

At present we do not observe baryon number violation, so the coupling
$a_{\mu}\bar{\psi}\gamma^{\mu}\psi$
between the baryon number current and the background field
can be eliminated by a field
redefinition $\psi \rightarrow \exp(i a_{\mu} x^{\mu}) \psi$.
However, for any experiments involving two species interacting
nontrivially (for example, the mass mixing between different
generation of quarks), the difference $\Delta a_\mu$ between the
corresponding two $a_\mu$ coefficients is observable. For the
quark sector, the mixing of neutral mesons provides an example
where this can be studied \cite{KP95, delta-a}. The experimental
constraint comes from the parameter
\be
  \Delta a_{\mu} = r_{q_1}
  a_{\mu}^{q_1} - r_{q_2} a_{\mu}^{q_2}\,,
\ee
where $a_{\mu}^{q_1}$,
$a_{\mu}^{q_2}$ are Lorentz-violating coupling constants for the
two valence quarks in the meson, and where the factors $r_{q_1}$
and $r_{q_2}$ allow for quark-binding or other normalization
effects \cite{KP95}. Experiments studying neutral $K$-mesons have
achieved sensitivity to two combinations of components of $\Delta
a$ involving the $a_\mu$ coefficients for $d$ and $s$ quarks, with bounds
in the Sun-centered frame of approximately
\be
  |\Delta a_{0}| \leq 10^{-20} {\rm ~GeV}
  \label{deltaa0}
\ee
by the KTeV Collaboration at Fermilab \cite{CPT2,KTeV}. Other
experiments with $D$ mesons have constrained two combinations of
$\Delta a$ for the $u$ and $c$ quarks at about $10^{-15}$ GeV (FOCUS
Collaboration, Fermilab) \cite{CPT2,FOCUS}.

Although there exists the possibility of a delicate cancellation
between the terms $r_{q_i} a_{\mu}^{q_i}$ contributing to
$\Delta a_{\mu}$, such an arrangement requires substantial
fine-tuning.  As we will see in Section~\ref{lorentz}, we require a
much larger expectation value $a_0$ than allowed by these constraints.
It is therefore necessary to invoke some mechanism by which the
value of $a_0$ changes substantially between the early universe and
today.

\subsection{Constraints from the axial vector current}

We consider the axial vector current for standard model fermions,
whose corresponding axial $U(1)$ symmetry is violated by the Dirac
mass term $m_{D}\bar{\psi^{c}} \psi$.
\begin{equation}\label{Eq8}
\mathcal{L}_{int} =  g \bar{\psi} \gamma^{\mu} \gamma^{5} \psi
A_{\mu} \rightarrow g a_{0} \bar{\psi} \gamma^{\mu} \gamma^{5}
\psi \, .
\end{equation}
If the earth is moving with respect to the  rest frame, then after
a Lorentz boost, it looks like the interaction $g \mu \bar{\psi}
\vec{\gamma} \gamma^{5} \psi \cdot \vec{v}_{earth}$. In the
non-relativistic limit, the current $\bar{\psi} \vec{\gamma}
\gamma^{5} \psi$ is identified with the spin density $s$, giving
us a direct coupling between the velocity of the earth and fermion
spin $g \mu \vec{s} \cdot \vec{v}_{earth}$.

Experimental limits on such couplings have placed considerable
bounds on $g a_{0}$. If we assume that the local  rest frame is
the same as the rest frame of the CMBR, then $|\vec{v}_{earth}|
\sim 10^{-3}$. The bound on couplings to electrons is $g a_{0}
\sim 10^{-25}$ GeV \cite{torsion} and to nucleons $g a_{0} \sim
10^{-24}$ GeV \cite{PAM00, C03}. These bounds put a strong limit
on the combination $g a_{0}$.

\subsection{Astrophysical constraints}

If the Lorentz-violating field arises as the gradient of a
scalar, $A_\mu = \partial_\mu\phi/f$, the chiral anomaly can induce
a coupling between $\phi$ and electromagnetism of the form
\begin{equation}\label{Eq13}
\mathcal{L}_{int} =  \frac{\gamma}{f} \phi F_{\mu\nu}\tilde{F}^{\mu\nu}
 \, .
\end{equation}
where $F^{\mu \nu}$ is the
electromagnetic field strength tensor and $\tilde{F}^{\mu \nu} =
{1\over 2}\epsilon^{\mu \nu \rho \sigma} F_{\rho \sigma}$ is its dual.
The dimensionless coupling $\gamma$ is typically of order
${e^{2}}/{4 \pi^2}$, and the dimensionful parameter $f$ sets the
scale of variation for $\phi$.  (It is,
however, possible to avoid this term entirely \cite{ACLM03}.)

A time-varying $\phi$ field would rotate the direction of
polarization of light from distant radio sources \cite{CF90}. The
dispersion relation for electromagnetic radiation in the presence
of a time-dependent $\phi$ becomes $\omega^2 = k^2 \pm (\gamma / f)
\dot{\phi} k$, where $+/-$ refer to right- and left-handed
circularly polarized modes, respectively. If we define $\chi$ to
be the angle between some fiducial direction in the plane of the
sky and the polarization vector from an astrophysical source, then
in the WKB limit where the wavelength of the radiation is much
less than that of $\phi$, the difference in group velocity for the
two modes leads to a rotation $\Delta \chi = \gamma(\Delta\phi)/f$.
If we can assume that $\dot\phi$ is roughly constant, this
becomes $\Delta \chi = \gamma{\dot\phi}(\Delta t)/f$,
where $\Delta t = t|_z - t|_{z = 0}$.

We use the data collected by Leahy \cite{Lea97}, the most stringent
bound on parameter $a_{0}$ comes from the single source 3C9 at $z
= 2.012$, which reads $\Delta \chi = 2^{\circ} \pm 3^{\circ}$, and
is consistent with the detailed analysis of Ref. \cite{WPC97}. This
gives us a tight bound on the parameter
\begin{equation}\label{Eq14}
 a_{0} \sim \frac{\dot\phi}{f}
 \leq \frac{\Delta \chi}{\gamma \Delta t} \sim 10^{-42}
 \textrm{GeV}\, .
\end{equation}
Although this constraint is extremely stringent, we take it somewhat
less seriously than (\ref{deltaa0}), since the interaction
(\ref{Eq13}) may be set to zero by an appropriate choice of dynamics.

\section{Sources of Lorentz violation}
\label{lorentz}

We now turn to the origin of the Lorentz-violating vector field
$A_\mu$.  We first examine the possibility that $A_\mu$
has a constant expectation value in the vacuum ($a_0 =$~constant).
This may arise either from fundamental vector (or higher-rank tensor)
fields, or as the gradient of a ghost condensate scalar field.  We
then consider slowly-rolling scalars, for which
the gradient $\partial_\mu\phi$ may not be constant.
In either case, we need to consider dynamics which allows for the
expectation value $a_0$ to be large enough to generate an appropriate
baryon asymmetry through (\ref{E16b}), while avoiding
constraints such as (\ref{deltaa0}).

\subsection{Fundamental vector fields}
\label{fundamental}

We consider a simple Lagrangian for a vector field $A_{\mu}$ and a
fermion $\psi(x)$.  For simplicity we choose the kinetic term for
the vector to be that of ordinary electrodynamics, although there
is no gauge invariance associated with the vector.  Nothing of
importance changes if we also include terms of the form
$(\nabla_\mu A^\mu)^2$ or $(\nabla_{(\mu}A_{\nu)})^2$; see
\cite{Lim:2004js,Jacobson:2004ts} for discussions of the gravitational
effects and
positive-energy degrees of freedom associated with such choices.

Along with the kinetic term and coupling to $\psi$, we introduce
a Mexican-hat potential for the vector field, so that it has a nonzero
vacuum expectation value.  The Lagrange density is thus
\begin{equation}\label{E5}
\mathcal{L} = -\frac{1}{4}F_{\mu \nu}F^{\mu \nu} + \bar{\psi} (i
\gamma \partial + m) \psi - g A_{\mu} \bar{\psi} \gamma^{\mu} \psi
- \frac{\mu^2}{2}A_{\mu}A^\mu - \frac{\lambda}{4}(A_{\mu}A^\mu)^2
 \, ,
\end{equation}
where $F_{\mu \nu} = \partial_{\mu} A_{\nu} - \partial_{\nu}
A_{\mu}$ denotes the field strength tensor for the vector field
$A_{\mu}$.
At the classical level, we obtain the minimal ground state energy
for a constant timelike vector field $A_{\mu}= (a_{0},
0, 0, 0)$, with the timelike component given by
\be
  {a_{0}}^2 = {\mu^2\over \lambda} .
\ee
Such vector-field condensation
leads to spontaneous breaking of Lorentz symmetry for a non-zero
time-like vev.

It is then straightforward to calculate the baryon-to-entropy ratio
produced by such a field.  In terms of the discussion in
Section~\ref{freeze-out}, we note that $\mu^0/T = ga_0/T$ is increasing
as the universe expands (and $T$ decreases).  The relevant freeze-out
temperature is thus $T_{+} = 150$~GeV, due to sphaleron transitions.
{}From eq.~(\ref{Eq-4}) we obtain
\begin{equation}\label{E9}
  {n_{B}\over s} \sim \frac{ga_{0}}{g_{\ast s}T_{+}}
  \sim ga_0 (10^4 {\rm ~GeV})^{-1}
 \, .
\end{equation}
If we plug in ${a_{0}}^2 = \mu^2/ \lambda$ and assume that both $\lambda$
and $g$ are of order unity,
obtaining the correct baryon density $n_B/s \sim 10^{-10}$
requires
\be
  \mu \sim 1 {\rm ~keV} .
\ee
The required Lorentz-violating effects are thus relatively small.
Thus, it is straightforward to obtain the correct baryon asymmetry from
a vector-field condensate that is constant throughout the electroweak
phase transition.  However, such a field would seem to violate the
experimental constraints discussed in Section~\ref{constraints},
unless there is some delicate cancellation that allows $A_\mu$ to
couple to the baryon current but avoid all other bounds.

One way to accomodate the experimental limits without fine-tuning
the interactions is to imagine a phase transition for the $A_\mu$ field
itself, which suddenly changes its expectation value at some point
after the electroweak scale.  Here we present one simple (albeit
contrived) example.

Consider the same Lagrangian of equation (\ref{E5}) but replace the
coefficient of the mass term $\mu^{2}$ with $({\mu'}^{2}- \alpha|\Phi|^2)$,
where $\Phi$ is the Higgs doublet:
\begin{equation}\label{Eqq1}
\mathcal{L} = -\frac{1}{4}F_{\mu \nu}F^{\mu \nu} + \bar{\psi} (i
\gamma \partial + m) \psi - g A_{\mu} \bar{\psi} \gamma^{\mu} \psi
- \frac{({\mu'}^{2}- \alpha|\Phi|^2)}{2}A_{\mu}^{2} -
\frac{\lambda}{4}(A_{\mu}^2)^2 \, ,
\end{equation}
At high temperatures, the Higgs expectation value $\langle\Phi\rangle$
vanishes, and we get a non-zero vacuum expectation value of the vector
background field. At late times, $|\Phi|^2 = v^2$. If $\mu'^2 -
\alpha v^2$ is negative, we get a zero vev for vector background field.
So the Lorentz symmetry is restored.  (Note that only the inequality
$\mu'^2 - \alpha v^2 < 0$ is required, not a strict equality.)

Another possibility, perhaps a more natural one, is
that finite-temperature effects lead to symmetry restoration
in the potential for $A_\mu$.  In other words, we imagine
that the coefficients $\mu$ and $\lambda$ are temperature-dependent,
such that the thermal corrections for the mass term are negative
and the expectation value $\langle A_\mu\rangle$ vanishes at zero
temperature.  We have not constructed an explicit model along
these lines, but this possibility deserves closer investigation.

We can also explore
more generic couplings containing high power of derivatives in the
current, as in Ref. \cite{BCKP96}.  Consider an interaction Lagrangian
\begin{equation}\label{E12}
\mathcal{L}_{int} \supset \frac{g \langle T
\rangle}{M^{k}}\bar{\psi} (\gamma^{0})^{k+1} (i
\partial_{0})^{k} \psi + h.c.
 \, ,
\end{equation}
where $\langle T \rangle$ is the vacuum expectation value of a
Lorentz tensor $T_{\mu_1 \cdots \mu_{k+1}}$
with dimension $k+1$, $g$ is a dimensionless
coupling constant, and $M$ is large UV cutoff mass scale.
In thermal equilibrium, we replace each time derivative with a
factor of the the associated fermion energy.  Due to the
existence of the high power of derivatives in the current, the
chemical potential is then no longer a constant; rather, it is
\begin{equation}\label{E13}
\mu \sim g \langle T \rangle \left(\frac{E}{M}\right)^k
 \, .
\end{equation}
We then get the baryon number density
\begin{equation}\label{E14}
\frac{n_{B}}{s} \sim g \frac{\langle T \rangle T_{F}^{k-1}
}{g_{\ast s} M^{k}}
 \, .
\end{equation}
If $k \geq 2$, the baryon number density is proportional to a
\textsl{positive} power of the freeze-out temperature $T_{F}$.

We therefore see that it is possible to get a phenomenologically
acceptable baryon asymmetry through a vector field with a constant
expectation value at temperatures at and above the electroweak
scale.  Of course, the need for a phase transition to subsequently eliminate
this expectation value renders the mechanism significantly less
attractive.

\subsection{Ghost condensates}
\label{ghost}

We now turn to the possibility that
the Lorentz-violating field arises as the gradient of a
scalar,
\be
  A_\mu = \frac{1}{f}\partial_\mu\phi\, ,
\ee with $f$ some parameter with units of mass. If the field
$\phi$ has an exact shift symmetry $\phi \rightarrow \phi + c$, it
always appears as derivatives in the action. Let us assume that it
has a wrong-sign quadratic kinetic term, i.e., it is a ghost field
\cite{ACLM03}.
\begin{equation}\label{Eq0}
\mathcal{L} = +\frac{1}{2}\partial_{\mu} \phi \partial^{\mu} \phi
+ \cdots
 \, .
\end{equation}
In the spirit of k-essence \cite{AMS00}, we assume that the
Lagrangian density has the general form $\mathcal{L} =P(X)$, where
$X =-\partial^{\mu} \phi \partial_{\mu} \phi$. The equation of
motion is
\begin{equation}\label{Eq1}
\partial_{\mu}[P'(X) \partial^{\mu} \phi]  = 0
 \, .
\end{equation}
{}From this we obtain a solution provided that $\partial_{\mu} \phi
= \textrm{constant}$.  In this case, then, the gradient
$\partial_{\mu}\phi$ behaves precisely like the constant vector field $A_\mu$ previously considered.

If $\partial_{\mu} \phi$ is time like, there
is a particular Lorentz frame where $\phi = - M^{2} t$. If we
consider small fluctuations about this ground state, the field
can be written
\be
  \phi = - M^{2}t + \pi\,,
\ee
where the $\pi$ field is a small fluctuation.
The kinetic energy is stable against small
excitations of $\pi$ provided that $P'(M^4) > 0$ and $P'(M^4) + 2
M^4 P''(M^4) > 0$ (so that the kinetic energy and
spatial gradient terms have the usual signs).

Then let us consider the direct couplings between the ghost field
$\phi$ and the Standard Model fields. The corresponding effective
Lagrangian density is
\begin{equation}\label{E1}
\mathcal{L}_{int} =  \frac{g}{f}\partial_{\mu} \phi J^{\mu}
 \, .
\end{equation}
(Note that this coupling violates the $\phi
\rightarrow -\phi$ symmetry, which is equivalent to violating
time-reversal invariance after the ghost field condenses.  Note also
that the leading interaction of the ghost condensate is typically to
axial vector currents, while our mechanism requires coupling to
the baryon or lepton vector currents.) After the ghost
field condenses so that $\phi \equiv - M^{2} t + \pi$, we have
\begin{equation}\label{E2}
\mathcal{L}_{int} = \frac{g}{f} ( - M^{2} J_{0} + J^{\mu}
\partial_{\mu} \pi )
 \, .
\end{equation}

If $J^{\mu}$ is the baryon number current, the first term in
Eq. (\ref{E2}) becomes $- \frac{g}{f} M^{2} (n_{b} -
n_{\bar{b}})$. The second term could be neglected not only because
it is small, but also the small fluctuating field $\pi$ is
oscillating around its minimum, which tends to cancel its
contributions \cite{DFRS97}. Plugging in $a_{0} = gM^2 /f$ in Eq.
(\ref{Eq-4}) we get:
\begin{equation}\label{E3}
n_{B}/s \sim \frac{gM^{2}}{f g_{\ast s} T_{F}}
 \, ,
\end{equation}
This scenario is very similar in spirit to that of the
fundamental vector field.
Once again, in order to avoid experimental bounds it is necessary
to have the ghost expectation value diminish dynamically between
the early universe and today.

\subsection{Quintessential baryogenesis}
\label{quintessential}

If the effective chemical
potential is time-dependent or, equivalently,
temperature-dependent, we will still get a nonzero net baryon
density.  A simple way to achieve this is to return to the possibility
that $A_\mu$ is the gradient of a scalar field $\phi$, but now
imagine that $\phi$ is a slowly-rolling quintessence field
\cite{WRP88, CDS98} rather
than a ghost condensate \cite{Trodden, XMZ, Yamag, DKKMS04}.
The chemical potential term $a_{0}$ is then given by $\dot{\phi}/
f$ \cite{Trodden, XMZ, Yamag}.  Quintessence models may be chosen
to have ``tracking'' behavior \cite{AS00}, in which
\be
\dot{\phi} \propto \sqrt{V(\phi)} \propto \sqrt{\rho_{\rm back}}\,.
\ee
In this case, during the radiation-dominated
era $\dot{\phi}$ is proportional to $T^{2}$.  We follow
the result of Ref. \cite{XMZ}, where the final net baryon density
is
\begin{equation}\label{Eqn6}
\frac{n_B}{s} \approx 0.01 g \frac{T_{F}}{f}
 \, ,
\end{equation}
where $f$ is the cut off scale for the effective coupling between
Lorentz-violating backgrounds and the baryon/lepton currents.
Since $\mu^{0} = {-g\dot{\phi}}/{f} \propto T^2$, according to Eq.
(\ref{E16a}), $T_{F} = T_{-}$ if there is $B-L$ violation in the
early universe. Another possibility is that there is no $B-L$
violation at all, so $T_{F}$ must be $T_{+} \approx$ 150~GeV. In
both cases, it is possible that the freeze-out temperature
$T_{F}$ could be around 1 TeV, and we can obtain the right baryon
number density with a cut off $f$ at the intermediate energy scale
around $10^{10}$~GeV.

A final model along these lines
\cite{DKKMS04} considers the interaction between the
derivative of the Ricci scalar curvature $\mathcal{R}$ and the
baryon number current $J^{\mu}$ from the effective theory of
gravity:
\begin{equation}
\frac{1}{M^{2}_{*}} \int d^{4}x \sqrt{-g} (\partial_{\mu}
\mathcal{R} ) J^{\mu} \, ,
\end{equation}
The net baryon number density obtained is proportional to an even
higher positive power of freeze-out temperature $T_{F}$, since
$\dot{\mathcal{R}} \propto \dot{\rho}$, where $\rho$ is the total
energy density.

\subsection{Baryogenesis from pseudo-Nambu-Goldstone bosons}
\label{pngbs}

If the Lorentz-violating background arises as the gradient of
a scalar field, one way to avoid equivalence-principle violating
Yukawa couplings is to imagine there is an
approximate shift symmetry $\phi \rightarrow \phi + {\rm constant}$
\cite{Carroll:1998zi}.
The field is then a  pseudo-Nambu-Goldstone boson (PNGB),
parametrized by a periodic
variable $\theta \sim \theta + 2 \pi$, which can be thought of
as the angular degree of freedom in a tilted Mexican hat potential.
Quintessential baryogenesis (and its relatives) imagines a field
that evolves gradually throughout the history of the universe,
perhaps with an energy density tracking that of the background.  In
contrast, a PNGB will remain overdamped in its
potential until the mass parameter becomes comparable to the
Hubble parameter, at which time it will roll to its minimum and
begin to oscillate.  During the initial rolling phase, the
gradient $\partial_\mu\phi$ acts like a Lorentz-violating vector
field, and leads to an intriguing scenario for baryogenesis.

We consider a generic
Lagrangian of the form \cite{FFO90}
\begin{equation}\label{Eqn1a}
\mathcal{L} = \frac{f^2}{2}(\partial \theta)^2 - \Lambda^4 [1-
\cos(\theta)]
 \, .
\end{equation}
The canonically normalized field is $\phi = f \theta$, and the
PNGB mass is
\be
  m = \frac{\Lambda^2}{f}\,.
\ee
Here, $f$ is the scale of spontaneous symmetry breaking giving
rise to the Mexican-hat potential, and $\Lambda$ is the scale
of explicit symmetry breaking that tilts the hat.
If we want to generate a baryon asymmetry of the right
amplitude, then from
\begin{equation}\label{Eqn1} \frac{n_{B}}{s} \sim
\frac{\dot{\phi}}{f g_{\ast s} T_{F}} = 10^{-10}
 \, ,
\end{equation}
with $g_{\ast s}\sim 100$ we require
\begin{equation}\label{Eqn2}
\dot{\phi} = 10^{-8} f T_{F}
 \, .
\end{equation}

The PNGB obeys the equation of motion
\be
  \ddot\phi + 3H\dot\phi + \frac{dV}{d\phi}=0\, .
\ee
If the field is oscillating near its minimum, $\dot{\phi}$ will
change sign quickly, cancelling any asymmetry \cite{DFRS97}.
We are therefore interested in the damped or overdamped regime,
in which we can ignore the second derivative term.  For typical
values $\phi \sim f$, we then have
\be
  \dot\phi \sim H^{-1}\frac{dV}{d\phi} \sim
  \frac{m^2 f M_{pl}}{T^2}\, ,
\ee
where $T$ is the temperature.  Thus, to achieve successful
baryogenesis requires that the freeze-out temperature satisfy
\be
  \frac{T_F^3}{m^2} \sim 10^8 M_{pl}\,.
  \label{pngbsb}
\ee

The ideal circumstance would be if freeze-out occured when the field
had just begun to roll substantially (becoming critically damped
rather than overdamped), but not yet begun to oscillate.  This
corresponds to $H\sim m$, which implies
\be
  T_F^2 \sim m M_{pl}\,.
\ee
Comparing to (\ref{pngbsb}) shows that PNGB baryogenesis works
if the freeze-out temperature is an intermediate scale
\be
  T_F \sim 10^{-8} M_{pl} \sim 10^{10}~{\rm GeV}
  \label{pngbfot}
\ee
and the PNGB mass is
\be
  m \sim \frac{T_F^2}{M_{pl}}
  \sim 100~{\rm GeV}\ .
\ee
The latter number is right around the electroweak scale,
which is an encouraging value for a moduli mass.  Note that the
condition $m\sim T_F^2/M_{pl}$ bears a resemblance to the PNGB
formula $m = \Lambda^2/f$; our conditions imply a model in which
$f$ is near the Planck scale, and
the explicit symmetry-breaking scale $\Lambda$ (perhaps arising
from strong dynamics) is at an intermediate scale around
$10^{10}$~GeV.  These requirements seem quite reasonable; in
particular, PNGB's with spontaneous-symmetry-braking scales
$f\sim M_{pl}$ arise naturally as axions in string theory.

This scenario depends on two separate conditions:
the PNGB must have a weak-scale mass, {\em and} the freeze-out
temperature must be (\ref{pngbfot}).
It is perhaps fair, however, to turn this around with a more
optimistic spin: if ($B-L$)-violating
interactions (so that the asymmetry is not diluted by sphalerons)
freeze out at an intermediate scale around $10^{10}$~GeV, a rolling
PNGB with electroweak-scale mass and a derivative coupling to
$J^\mu_{B-L}$ will naturally give rise to a baryon asymmetry of the
correct magnitude.  In fact it is quite natural to imagine
($B-L$)-violating interactions that freeze out around this scale,
arising for example from Majorana neutrino masses.  This scenario
seems to be worthy of further investigation.

\section{Discussion}

We have investigated the possible origin of the observed baryon
asymmetry in the presence of a coupling between a Lorentz-violating
vector field and the baryon current.  In the right circumstances
it is possible to generate the asymmetry without
any baryon-violating interaction beyond the Standard Model, as
sphaleron transitions will violate baryon number and
drive a matter and antimatter asymmetry in thermal
equilibrium.  Baryon violation that freezes out at the weak
scale can also be converted to an appropriate asymmetry by
the evolution of pseudo-Nambu-Goldstone bosons with weak-scale
masses.  The generic conditions and our comments on
such kind of baryogenesis are summarized as follows:
\begin {itemize}
\item
We need a background field with a nonzero time-like vev which
spontaneously breaks Lorentz invariance in the early universe.
The vev is not necessarily a constant, but can vary with time or
even change discontinuously through phase transitions.
\item
Such background field may couple to the baryon number
current $J_{B}^{\mu}$ or the lepton number current $J_L^{\mu}$,
as sphalerons violate $B+L$ and can turn a lepton asymmetry into
a baryon asymmetry.  In the absence of any symmetry, the couplings
$g_{-} A_{\mu} J_{B-L}^{\mu}$ and $g_{+} A_{\mu} J_{B+L}^{\mu}$
should be of similar magnitude.
\item
We have considered baryon/lepton-violating process in thermal
equilibrium. If $a_{0}/T$ is increasing with time, then the
final net baryon number density is determined by the freeze-out
temperature $T_{+} \approx 150 $~GeV. For the opposite case, the
final net baryon number density is determined by the freeze-out
temperature $T_{-}$ which is model-dependent.
\end{itemize}

Our discussions and conclusions here differ from the
usually baryogenesis mechanism in thermal equilibrium. In most
previous works \cite{BCKP96, Trodden, XMZ, Yamag, DKKMS04}, the
absolute value of the effective chemical is decreasing with time,
so that in order to generate a right net baryon number density, we
need a high freeze-out temperature $T_{-}$ as $n_{B-L}/s$ depends
on the positive power of $T_{-}$. However, in our paper, the
Lorentz-violating background can be a constant vev instead of a slow
rolling field, so that sphaleron transitions
can be the main source to generate the baryon asymmetry and the
energy scale for baryogenesis to happen is low.

Finally, we want to discuss some generic features of the
background fields here for a successful baryogenesis. We have seen
that in order to obtain the right net baryon number density in the
early universe, the coupling times the time component of the
background field $g a_{0}$ should be not too small. However, such
spontaneous Lorentz-violating term is highly constrained by the
experiment at present, so we need some dynamical mechanism to
decrease the time component value of such background field. If the
background field is some vector or tensor field, we may imagine
that a phase
transition occurs in between freeze-out and today, so as to evade
the experimental constraint at present. If the background arises
as the derivative of a scalar, two possibilities present themselves:
the field could roll gradually in a tracking-type potential with
a gradually diminishing value of $\dot\phi$, or it could roll
only temporarily before reaching the bottom of its potential and
oscillating.  The former possibility leads to quintessential
baryogenesis, while the latter is realized in
pseudo-Nambu-Goldstone boson models.

Our investigation of the PNGB scenario reveals that the most
natural implementation of this idea requires PNGB's with
weak-scale masses (100~GeV) and $(B-L)$-violating interactions
that freeze out at an intermediate scale of around
$10^{10}$~GeV.  The former condition can be satisfied if the
spontaneous symmetry-breaking scale $f$ is of order the Planck
scale and the explicit symmetry-breaking scale $\Lambda$ is at
the same intermediate scale, while the latter condition arises
naturally from the decay of Majorana neutrinos.  We therefore
consider this scenario to be quite promising.

At least two extensions of our investigation remain for future
work. One is that we have only considered baryon/lepton number
violation. As we have mentioned in footnote 2, any currents
derivatively coupled to the scalar field that are non-orthogonal to
the baryon number current could possibly lead to a non-zero net
baryon number. A simple case is that the scalar field derivatively
couples to a current associated with global symmetry $U(1)_Q$, and
an interaction violates baryon number and $U(1)_Q$
simultaneously \cite{Chiba:2003vp}. More complicated cases should
be considered, especially when the baryon-number violation freezes
out around the electroweak phase transition. The other case is to
consider baryon number violation freeze-out when scalar $\phi$ is
in the oscillation stage, as in spontaneous baryogenesis.

\section*{Acknowledgements}  We would like to thank
Alan Kostelecky, Joseph Lykken, David Morrisey, Mark Trodden, and
Carlos Wagner for helpful conversations.  This work was supported
in part by the U.S. Dept. of Energy, the National Science
Foundation, and the David and Lucile Packard Foundation.  The KICP
is an NSF Physics Frontier Center.

\section*{Appendix: single-particle Hamiltonian}

In this Appendix we verify that the effect of the coupling of
fermions to a Lorentz-violating vector field is to induce an
effective chemical potential $\mu^0$, by evaluating the
Hamiltonian for a point particle in such a background.

We consider a particle with interaction Lagrangian density
(\ref{Eq-2}). Suppose we have a system of fermions with positions
$x_{n}(t)$ and net conserved fermion number $Q$. The current
$J^{\mu}$ for a single particle in special relativity is \cite{Wein}
\begin{equation}\label{Eq2-3a}
J^{\mu} \equiv \sum_{n}Q_{n}\delta^3 (x-x_{n}(t))\frac{d
x_{n}^{\mu}(t)}{d t}
 \, .
\end{equation}
Integrating over space, the new interaction is
\begin{equation}\label{Eq2-3b}
L_{int} = -gA_{\mu}\tilde{Q}\frac{dx^{\mu}}{d t}
 \, ,
\end{equation}
where $\tilde{Q}$ is the conserved charge per particle. The
canonical momentum $P$ conjugate to the position coordinate $x$ is
given by
\begin{equation}\label{Eq2-3d}
P_{\mu} \equiv \frac{\partial L}{\partial x^{i}} = \frac{\partial
L_{matter}}{\partial x^{i}} + \frac{\partial L_{int}}{\partial
x^{i}} = p_{i} + g A_{i} \tilde{Q}
 \, ,
\end{equation}
where $p_{i} = \gamma m d x_{i} / d t $ is the ordinary kinetic
momentum from the fermion sector and $i= 1,2,3$
runs over the spatial components. The corresponding Hamiltonian
is
\begin{eqnarray}\label{Eq2-3c}
H &=& p_{i} \frac{d x^{i}}{d t} - L_{int} - L_{matter} \nonumber
\\ &=& (p_{i} + gA_{i} \tilde{Q} )
\frac{d x^{i}}{d t} + g\tilde{Q}A_{0} - L_{matter}
\nonumber \\ &=& P_{i} \frac{d x^{i}}{d t} - L_{matter} +
g\tilde{Q}A_{0} \, .
\end{eqnarray}
If we solve for $d x_{i} / d t $ in terms of canonical momentum
and plug it into the Hamiltonian, we will find the Hamiltonian for
relativistic fermions coupled to a vector field. (A similar
discussion to the physics here could be found in Chapter 12.1 of
Ref. \cite{e&m}.)
\begin{equation}\label{Eq2-6}
{H} = \sqrt{({P_{i}} - g\tilde{Q} A_{i})^2 + m^2} + g\tilde{Q}
A_{0}
 \, .
\end{equation}
Note that the above result does not depend on the fact whether the
vector background field has gauge invariance or not.

The vector background field picks out a nonzero
constant vev at the ground state,
$\langle A_{\mu}\rangle = (a_{0}, 0, 0, 0)$.
The interaction Lagrangian density becomes
\begin{equation}\label{Eq2-4}
\mathcal{L}_{int} = - g a_{0} \tilde{Q}
 \, .
\end{equation}
The Hamiltonian for one particle is
\begin{equation}\label{Eq2-7}
{H} = \sqrt{{P_i}^2 + m^2} + g\tilde{Q} a_{0} = \sqrt{{p_i}^2 +
m^2} + g\tilde{Q} a_{0}
 \, .
\end{equation}
We see that the additional term $gQ a_{0}$ acts like a
``potential" term for the relativistic charged particle. In the
statistics of relativistic charged fermions, such potential term
$gQa_{0} = Ng\tilde{Q}a_{0} \propto N$, so it can be
considered as a contribution to the chemical potential. We define
the ``effective chemical potential"
\be
\mu^{0} \equiv g \tilde{Q} a_{0} \ .
\ee
If we have interactions that violate the charge $Q$ (such as
baryon number $B$), then $\sqrt{{p_i}^2 + m^2}$ will not be a
conserved number due to energy conservation. This will lead to an
asymmetric distribution of the particles and generate a final
non-zero net charge $Q_{f}$.

\end{document}